\documentclass{revtex4}

\usepackage{graphicx}
\usepackage{amssymb,bm,amsmath}
\usepackage[dvips]{color}

\font\amsb=msbm10
\def\hbar{\mbox{\amsb\char'175}}

\newcommand{\be}{\begin{equation}}
\newcommand{\ee}{\end{equation}}

\newcommand{\vxi}{{\mathbf \xi}}
\newcommand{\x}{{\mathbf x}}
\newcommand{\X}{{\mathbf X}}

\newcommand{\y}{{\mathbf y}}

\newcommand{\J}{{\mathbf J}}

\newcommand{\vct}[1]{\ensuremath\mbox{\boldmath$ #1 $}}
\newcommand{\Vxi}{\vct \xi}
\newcommand{\Veta}{\vct \eta}

\def\eq{{\xi_q}}
\def\ep{{\xi_p}}

\newcommand{\braket}[2]{\ensuremath{\left\langle#1\big|#2\right\rangle}}

\newcommand{\parcial}[2]{\ensuremath{\frac{\partial#1}{\partial #2}}}

\newcommand{\dparcial}[3]{\ensuremath{\frac{\partial^{2}#1}{\partial#2\partial#3}}}
\newcommand{\deri}[2]{\ensuremath{\frac{d #1}{d #2}}}

\def\Ai{{\rm Ai}}

\begin{document}

\title{Uniform approximation for the overlap caustic
of a quantum state with its translations}

\author {Eduardo Zambrano\footnote{zambrano@cbpf.br} and
         Alfredo M. Ozorio de Almeida\footnote{ozorio@cbpf.br}}
\address{Centro Brasileiro de Pesquisas F\'\i sicas,
Rua Xavier Sigaud 150, 22290-180, Rio de Janeiro, R.J., Brazil}

\begin{abstract}

The semiclassical Wigner function for a Bohr-quantized energy eigenstate
is known to have a caustic along the corresponding classical closed 
phase space curve in the case of a single degree of freedom. 
Its Fourier transform, the semiclassical chord function,
also has a caustic along the conjugate curve defined as the locus of diameters,
i.e. the maximal chords of the original curve. If the latter is convex, 
so is its conjugate, resulting in a simple fold caustic.
The uniform approximation through this caustic, that is here derived,
describes the transition undergone by the overlap of the state with its translation,
from an oscillatory regime for small chords, to evanescent overlaps,
rising to a maximum near the caustic. The diameter-caustic for 
the Wigner function is also treated.

\end{abstract}

\maketitle

\section{Introduction}

It is often assumed that the Wigner function \cite{Wigner},
$W(\x)$ in the phase plane $\x=(p, q)$,
for a semiclassical WKB state can be approximated
by a Dirac $\delta$-function on the corresponding
classical closed curve in phase space. This simplest
approximation does, indeed, lead to reliable expectation values
for smooth classical observables. Nonetheless,
it is hopelessly inadequate for the description
of delicate interference effects that are becoming
ever more accessible to experiments related to
quantum information, either in quantum optics,
atom traps, or other quickly developing technologies.
It is then necessary to resort to more refined
semiclassical descriptions for the phase space representations
of quantum states, such as Berry's uniform approximation
for the Wigner function \cite{Berry77}.

A typical interference experiment superposes two modified copies
of the same initial state (see e.g. \cite{Leonhardt}). 
For instance, in quantum optics, it is easy to achieve
the unitary transformation that corresponds to a uniform
phase space translation (or displacement). 
This translated state can then interfere with the original state.
In general, the unitary {\it translation operator}
\begin{equation}
\hat{T}_{\Vxi} = \exp{\left[\frac{i}{\hbar}(\Vxi\wedge \hat{\x})\right]}=
\exp{\left[\frac{i}{\hbar}(\Vxi_p\cdot \hat{q}-\Vxi_q\cdot \hat{p})\right]} \ ,
\end{equation}
acts on the state $|\psi\rangle$ to produce the new state
$|\psi_{\Vxi}\rangle=\hat{T}_{\Vxi}|\psi\rangle$ in strict
correspondence to the classical translation: $\x \mapsto \x + \Vxi$.
\footnote{ In the optical context $\hat{T}_{\Vxi}$ is
usually referred to as the {\it displacement operator} and is
expressed in terms of creation and annihilation operators for the
harmonic oscillator. This is inconvenient for semiclassical
analysis.}
Thus, given an arbitrary superposition of a state and its translation,
$a|\psi\rangle+b|\psi_{\Vxi}\rangle$, with $|a|^2+|b|^2=1$,
the probability that this is measured to be in the untranslated state
is $|a+ b\langle\psi|\psi_{\Vxi}\rangle|^2$.

Evidently, measurements of such probabilities (through
repeated preparation) supply detailed quantum information
concerning these initial states.
It so happens that the full set of possible overlaps defines the
complete phase space representation,
\be
\chi (\Vxi)= \frac1{(2\pi\hbar)}\langle\psi|\hat{T}_{-\Vxi}|\psi\rangle.
\label{chitrans}
\ee
This is known as the {\it chord function} \cite{OzorioRep}, 
the {\it quantum characteristic function}
(or the {\it Weyl function} as in \cite{Choun}),
which is the Fourier transform of the Wigner function:
\begin{equation}
 \chi(\Vxi) = \frac{1}{(2\pi\hbar)}\int d\x \;W(\x)
\exp{\left\{\frac{i}{\hbar}(\Vxi\wedge \x)\right\}} \ .
\end{equation}
The latter can be redefined, following Royer \cite{OzorioRep,Royer}, as
\be
W(\x)=\frac{1}{(\pi\hbar)}\langle\psi|\hat{R}_{\x}|\psi\rangle,
\label{Wrefl}
\ee
where $\hat{R}_{\x}$, the Fourier transform of the translation operators,
corresponds classically to the phase space reflection through the point $\x$,
i. e. $\x_0 \mapsto 2\x - \x_0$. An important consequence is that the phase space correlations
\cite{Choun, OzVaSa}, for translational interference,
\begin{equation}
 C(\vxi)=|\langle\psi|\psi_{\Vxi}\rangle|^2=
(2\pi \hbar)^2  \int \rm d\Veta \; e^{ i \Veta \wedge \Vxi / \hbar}
\left| \chi(\Veta )\right|^2=
(2\pi \hbar)^2 \int \rm d\x \>W(\x)\> W(\x-\Vxi) \; ,
\label{pscor}
\end{equation}
coincide with the autocorrelation of the Wigner function itself. 
A study of interference phenomena using the quantum phase space formalism, 
besides some properties of phase space distributions can be found in \cite{Choun}.

In the limit of small displacements, $\Vxi\rightarrow 0$, the
correlations attain their maximal value, $C(0)=1$. Increased
translations reduce them to an oscillatory regime. We shall have
be concerned with the correlation of states, whose chord function
can be semiclassically approximated by \cite{OzVaSa, Oz-enta}
\begin{equation}
\label{semichord} \chi(\Vxi) = \sum_j \;
\alpha_j(\Vxi)\;e^{i{\sigma}_j(\Vxi)/\hbar} = \sum_j \;
\chi_j(\Vxi)\ ,
\end{equation}
where the amplitudes and phases are determined by a classical curve,
as will be described in the next section. This is
similar to the simplest semiclassical approximation for the
Wigner function \cite{Berry77}
\begin{equation}
\label{semiwigner}
W(\x) = \sum_{j}\; a_j(\x) \;e^{i
A_j(\x)/\hbar} = \sum_j \; W_j(\x) \ .
\end{equation}
Furthermore, the Fourier relation between this pair of
representations is reflected in the reciprocal relation, which
specifies the {\it centres}
\begin{equation}
\x_j(\Vxi) =  \J \frac{\partial \sigma_j}{\partial \Vxi} \ ,
\end{equation}
for the realizations of the vector $\Vxi$ as a {\it chord}
of the classical curve, whereas
\begin{equation}
\Vxi_j(\x) =  -\J \frac{\partial S_j}{\partial \x} \ .
\label{chorderiv}
\end{equation}
determines the chords that have a given centre.
Here,
\begin{equation}
\label{eqc.4}
\J=\left(
\begin{array}{cc}
    0 & -1 \\
    1 &  0
\end{array}
\right) \;
\end{equation}
is the standard symplectic matrix. Typically for an energy
eigenstate, the classical curve is a level curve for the
corresponding classical Hamiltonian, related by Bohr-Sommerfeld 
\cite{VanVleck, Maslov, OzorioHam, Gutzwiller} quantization.  
In this way \eqref{semichord} and \eqref{semiwigner} are
alternative phase space representation of WKB wave functions.

For large enough  displacements, such that
the classical translation of the curve does not intersect the original curve,
the phase space correlations are negligible.
The transition between this and the previous oscillatory regime
takes place along a caustic where the chord function attains
locally maximal amplitudes and the simple semiclassical approximation
(\ref{semichord}) breaks down. The main purpose of this paper is to
establish the correct description of this transition from the oscillatory regime
to the region of negligible overlap through a uniform approximation. 
It is interesting that the caustic region for increased quantum correlations 
is entirely determined by the geometry of the classical curve
supporting the quantum state. It will here be assumed
that this curve is convex, so that the locus of its diameters,
i.e. its maximal chords, also defines a closed convex curve.

The present approximation has some resemblance to the uniform
approximation obtained by Berry \cite{Berry77} for the Wigner
function close to the classical curve. However, the latter is
simplified by symmetry constraints that do not hold here. Indeed,
the present treatment is even closer to the uniform approximation
along the caustic of the Wigner function far from the curve, which
will also be treated. For a start, section 2 reviews the
geometrical construction of the Wigner function and the chord
function for a state that corresponds to a closed quantized curve.
In contrast to the treatment in \cite{Berry77}, the construction
of the present uniform approximations cannot be limited to a
single WKB branch.

Having defined geometrically the stationary phases for the
Wigner function and the chord function, the method of Chester,
Friedman and Ursell \cite{Chester} then supplies uniform approximations
for the chord function in terms of the Airy function and its
derivative in section 3 and for the Wigner
function in section 4. In both cases, the asymptotic form of these functions
for large argument are then connected to the simpler semiclassical
forms (\ref{semichord}) and (\ref{semiwigner}).
The analysis of these uniform approximations close to the caustic
furnishes simpler transitional approximations in section 5.

None of these expressions extends right  down to the limit of
small chords, because this is another caustic for both the Wigner
function and the chord function. However, for the eigenstates of
the harmonic oscillator (Fock states), the approximations for
large chords can be compared to a small chord formula specified by
a Bessel function\cite{OzVaSa}. This leads to a discussion in
section 6 of the normalization of all these phase space
approximations, deriving from the limit $C(0)=1$.

\section{Construction of the semiclassical Wigner and chord functions}

Before embarking on uniform approximations for the Wigner and chord functions,
it is worthwhile to review the derivation of the simpler semiclassical
formulae presented in the introduction. Thus we also specify explicitly
the amplitudes and phases and their geometric interpretation, which are
essential ingredients of the uniform approximations.

The starting point is the generalized WKB expression for the wave function
\cite{VanVleck, Maslov, OzorioHam, Gutzwiller},
\begin{equation}
\label{WKB}
\braket{q}{\psi_I}=
                    N\sum_j\left|
                                 \dparcial{S_j(q,I)}{q}{I}
                           \right|^{\frac12}
                    \exp\left[
                              \frac{i}\hbar S_j(q,I)+i\beta_j
                       \right].
\end{equation}
Here we assume that the classical curve is defined as a level curve
of the action variable $I(\x)$, such that $S_j(q,I)$ corresponds to
the $j$'th branch of the generating function for the canonical
transformation $(p,q)\mapsto(I,\theta)$, $\beta_j$ is the Maslov
correction and $N$ is the overall normalization constant. 
Choosing the arbitrary initial point, $q_0$, the action branches
are defined as
\begin{equation}
S_j(q,I)=\int_{q_0}^{q} p_j(Q,I)\; dQ,
\quad
\quad
\parcial{S_j}{I}=\theta_j,
\quad
\quad
\parcial{S_j}{q}=p_j,
\end{equation}
so that the amplitude can be rewritten in terms of
\begin{equation}
\dparcial{S_j(q,I)}{q}{I}=
\parcial{p_j}{I}(q)=
\left[
      \parcial{I}{p}(p_j(q),q)
\right]^{-1}.
\end{equation}

It will be assumed that the state corresponds classically
to a convex closed curve, so that there will always be a single pair of
branches for the action function. This will hold irrespective of
any linear canonical transformation, which it may be convenient to
make, given that both the chord function and the Wigner functions
are covariant with respect to such changes of phase space
coordinates. It will also be important to recall that the closed
curve must satisfy the Bohr-Sommerfeld quantization condition:
\begin{equation}
\oint p\,dq = 2\pi\hbar\left(n+\frac12\right),\quad\quad\quad n\in\mathbb{Z}.
\label{Bohr}
\end{equation}

Expressing the translation and reflection operators within
the position representation (see e.g. \cite{OzorioRep}),
the chord function \eqref{chitrans} becomes
\begin{equation}
\label{chi}
\chi(\xi)=\frac1{(2\pi\hbar)}\int dq
\braket{q^+}{\psi_I}\braket{\psi_I}{q^-}e^{-i\bm{\xi}_p\cdot q/\hbar}
\end{equation}
while the Wigner \eqref{Wrefl} is given by
\begin{equation}
\label{wigner}
W(\x)=\frac1{(\pi\hbar)}\int d\bm{\xi}_q
\braket{q^+}{\psi_I}\braket{\psi_I}{q^-}e^{-ip\cdot\bm{\xi}_q/\hbar}
\end{equation}
where, in both equations $q^\pm=q\pm\bm\xi_q/2$.

\begin{figure}[t!]
\centering
\includegraphics[width=7cm]{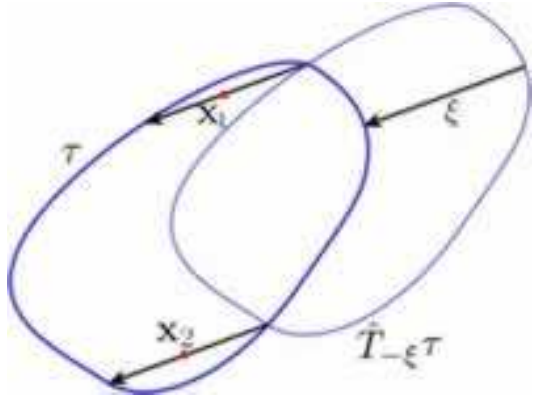}
\caption{Geometrical construction for the stationary points of the chord function.
The intersections at ${\x_1}^-$ and ${\x_2}^-$, of the closed curve
with its translation by $-\Vxi$, define the pair of realizations
of this chord. The $q$-coordinate of the centres,
${\x_1}$ and $\x_2$ of these realizations defines the stationary points.
\label{p.e.corda}}
\end{figure}

In the semiclassical limit, the WKB expression can be inserted,
so that we obtain in each case a sum of integrals that are dominated
by their points of stationary phase. Irrespective of whether these
points are sufficiently isolated  so as to allow for immediate evaluation
by the stationary phase method, we need to understand the geometric
construction that defines them. In both cases, each stationary point
defines  values of $q^\pm$ pairs, which are $q$-coordinates
of a pair of points, $\x^\pm$, lying on the classical closed curve.
In the case of the chord function, each $\x^-$ is the intersection
of the classical curve with its uniform translation by the vector $-\Vxi$,
whereas $\x^+={\x^-}+\Vxi$. This geometry is exhibited in Fig. \ref{p.e.corda},
which shows that each chord has two {\it realizations} in a convex
closed curve. Thus, this construction on a given convex curve always
specifies a pair of centres $\x_1(\Vxi)$ and $\x_2(\Vxi)$ for each
chord $\Vxi$ that can be fitted in the curve.

\begin{figure}[t!]
\centering
\includegraphics[width=7cm]{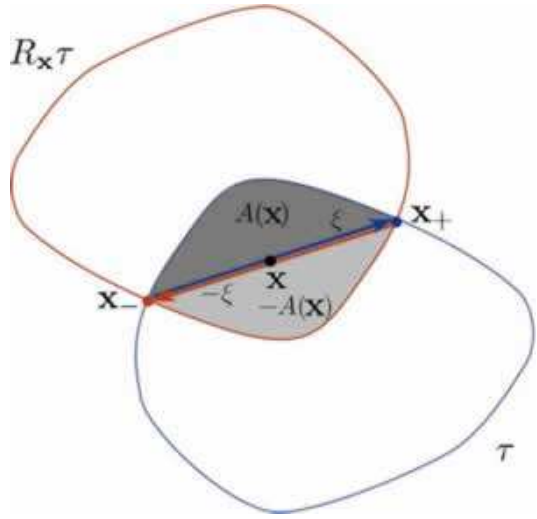}
\caption{Geometrical construction for the stationary points of the Wigner function.
The intersections of the closed curve with its reflection through $\x$
defines pair of chords centred on $\x$. The $q$-coordinate of these
chords defines the stationary points.
The area $A(\x)$ determines the phase of the Wigner function.}
\label{A-wigner}
\end{figure}

In the case of the Wigner function, instead of a translation,
the classical curve is reflected through the {\it reflection centre},
$\x$.  This results in a pair of intersections, $\x^+$ and $\x^-$, such that $\x$
is the centre for the pair of chords $\Vxi= \pm(\x^+ - \x^-)$.
The stationary points are then the $q$-coordinates for this pair of chords.
This geometry is shown in Fig. \ref{A-wigner}. Therefore, there will be at
least one pair of chords $\pm\Vxi(\x)$, for each reflection centre in the curve.

The stationary phase evaluation of each integral for the Wigner function,
\begin{equation}
W_{ij}(\x)=
\frac{N^2}{2\pi\hbar}\hspace{-.1cm}
\int \hspace{-.1cm} d\bm{\xi}_q\hspace{-.1cm}
                \left|
                      \parcial{ I}{p}(p_i(q^+),q^+)
                      \parcial{I}{p}(p_j(q^-),q^-)
                \right|^{\frac12}\hspace{-.12cm}
\exp\left[
          \frac{i}\hbar\left[S_i(q^+,I)-S_j(q^-,I)-p
          \cdot\hspace{-.05cm}\hspace{-.05cm}\bm{\xi}_q+\beta_i-\beta_j\right]
     \right]\hspace{-.1cm},
\label{WigKB}
\end{equation}
can usually be obtained from a single branch of the action function
for the closed curve, that is, $i=j$. The stationary phase is half of the area between
the curve and its reflection, or the ``chord area" as shown in Fig. \ref{A-wigner},
except for a Maslov correction. The only difference between both phases, 
corresponding to $\pm\Vxi(\x)$, is the sign, 
so that the semiclassical approximation is real \cite{Berry77}.

\begin{figure}[b!]
\centering
\includegraphics[height=4cm]{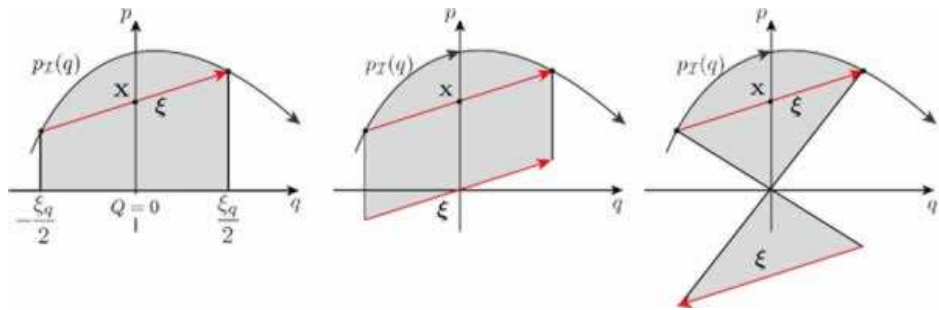}
\caption{Several geometrical interpretations for the phase for the semiclassical
chord function in the simpler approximation for a WKB function, considering one
branch. The stationary phase itself is determined 
by the shaded area in the three cases\cite{OzVaSa}.
\label{ph-chordOzVaSa}}
\end{figure}

In the case of the chord function, each stationary phase is given by the
construction in Fig. \ref{ph-chordOzVaSa}.
Unlike the Wigner function, this depends explicitly
on a change of phase space origin, $\x\mapsto\x+\Vxi'$,
according to the exact formula \cite{OzVaSa},
\begin{equation}
\label{transchord} \chi_{\vct \Vxi'}(\Vxi) =
e^{\frac{i}{\hbar}\vct \Vxi'\wedge\Vxi}\; \chi(\Vxi)  ,
\end{equation}
but it is also covariant with respect to homogeneous linear
canonical transformations.

The amplitudes in the above semiclassical
approximations are best expressed in terms of
the canonical action variable, $I(\x)$, that defines
the closed curve and is conjugate to the angle variable $\theta(\x)$ along the curve.
If we now define the transported action variable,
\be
I^\pm=I(\x\pm \Vxi/2),
\label{transaction1}
\ee
then, generally, the Poisson bracket for this pair of functions,
\be
\{I^+, I^-\}=\parcial{I^+}{\x}\wedge\parcial{I_-}{\x}\neq0,
\ee
and it is found that these amplitudes in \eqref{semichord} and \eqref{semiwigner} are
\be
 a(\x)= | \{I^+, I^- \}|^{-\frac12} =\alpha(\Vxi).
\label{amplitude} \ee The difference between considering the
amplitude as a function on $\x$ or $\Vxi$ depends on which
variable is fixed in \eqref{amplitude}. This only changes the sign
of the Poisson bracket. Here, the equality of coefficients for
both representations holds for the chord and its centre, between a
specific pair of points $(\x^-,\x^+)$ on the closed curve.

\begin{figure}[t!]
\centering
\includegraphics[width=8cm]{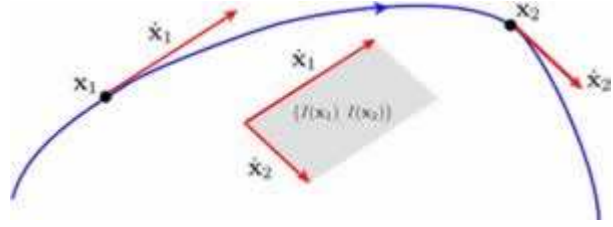}
\caption{Geometrical interpretation for the semiclassical amplitudes
of both the Wigner function and the chord function.}
\label{colchete}
\end{figure}

A neat interpretation for these amplitudes follows from the identification
of the action variable $I(\x)$ with a classical Hamiltonian.
Then the closed curve becomes a closed trajectory, tangent to
the phase space velocity vector, $\dot{\x}$, and
\begin{equation}
\label{colchete-x x}
\{I^-,I^+\}=\dot\x^-\wedge\dot\x^+,
\end{equation}
as shown in Fig. \ref{colchete}.
It follows that the amplitudes
$\alpha_j(\Vxi)$ (or $a_j(\x)$), depend on the degree of
transversality of the intersection between
the curve and its translation, or its reflection \cite{Berry77, OzVaSa}
and so they diverge at caustics, where $\dot\x^+$ and $\dot\x^-$ are parallel.

So far, we have presumed that the contribution of each stationary
phase point to the integrals for either the Wigner function or the
chord function, can be obtained by considering a single branch of
the WKB wave function. Thus, $i=j$ in \eqref{WigKB} and this single
branch {can always be} accessed through a canonical phase space rotation in the
simple semiclassical approximations above and even for Berry's
uniform theory for the caustic that arises in the limit of small
chords \cite{Berry77}. However, this is not possible in the
present treatment of the caustic at maximal chords, where
\eqref{amplitude} also diverges for both representations, because
the tips of the stationary chord become turning points for the WKB
function in this limit. It is thus necessary to study this limit
with the aid of phase space coordinates, such that it is the cross
term between the pair of different branches, of the action function, 
which have stationary phases in \eqref{WigKB}. 
For this reason it will be important to take care of
the phase relation between the branches across a turning point:
\begin{equation}%\label{WKB-2}
\braket{q}{\psi_I}=
N\left[\left|\dparcial{S_+(q,I)}{q}{I}\right|^{\frac12}
                    \exp\left[\frac{i}\hbar S_+(q,I)\right]
+{\rm e}^{i\pi/2}
   \;\left|\dparcial{S_-(q,I)}{q}{I}
                           \right|^{\frac12}
                    \exp\left[
                              \frac{i}\hbar S_-(q,I)\right]\right],
\label{WKBwf}
\end{equation}
where $\pi/2$ is the Maslov phase and $N$ is the normalization constant. 
Inserting the above wave function in \eqref{chi}, we obtain four integrals to evaluate. 
An appropiate choice of the coordinate axes orientation,
in which the both of chord realizations are crossed,  
reduces the chord function for \eqref{WKBwf} to the single integral
\begin{eqnarray}
\chi(\Vxi)&=&\frac{N^2}{2\pi\hbar}
\int dq
                \left|
                      \parcial{ I}{p}(p_+(q^+),q^+)
                      \parcial{I}{p}(p_-(q^-),q^-)
                \right|^{\frac12}\hspace{-.1cm}
\nonumber
\\
&&\hspace{1cm}\times\exp\left(
          \frac{i}\hbar\left[S_+(q^+,I)-S_-(q^-,I)-\bm{\xi}_p\cdot q\right]+i\frac\pi2
     \right).
\label{chiKB}
\end{eqnarray}
Here $q^\pm=q\pm\Vxi_q/2$.  As stated previously, 
this geometry can be guaranteed by a phase space rotation.
The evaluation of the chord areas in the Wigner function and the chord function 
for this geometry is discussed in Appendix A.

%%%%%%%%%%%%%%%%%%%
\section{Uniform approximation for the chord function}

Let us allow the pair of these stationary points of \eqref{chiKB} to coalesce
for the chord $\Vxi_D$, that corresponds to a {\it diameter} of the closed curve,
i.e. a maximal chord, at which the semiclassical amplitude \eqref{amplitude} diverges.
In the present case of a convex curve, these diameters are the locus of a fold caustic
with no higher singularity. This is simpler than the geometry for the corresponding
caustic of the Wigner function, studied in the following section.

\begin{figure}[t!]
\centering
\includegraphics[width=5cm]{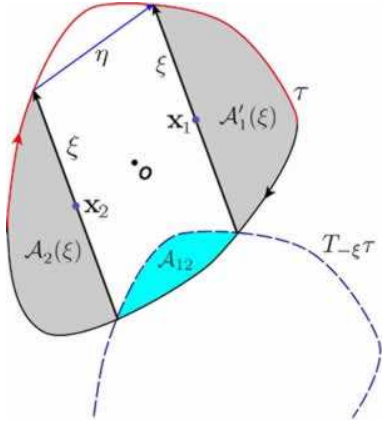}
\caption{The difference in chord areas, $\mathcal{A}_{12}$,
coincides with the area between the closed curve and its
translation. This area shrinks to zero when $\Vxi$ becomes a
diameter and its conjugate chord $\eta=\x_1-\x_2\rightarrow 0$.
The simplest choice of phase space origin is at midpoint between
$\x_1$ and $\x_2$.} \label{area-corda}
\end{figure}

We also simplify the calculation by an appropriate
choice of origin, in view of the simple translation property
\eqref{transchord} of the chord function. This ideal origin
lies midway between the centres for the pair of chord realizations, shown in Fig. \ref{area-corda}.
The pair of stationary points $q_1$ and $q_2$ of the integrand
in \eqref{chiKB} are solutions of the equation:
\begin{equation}
p_+(q+\eq/2)-p_-(q-\eq/2)=\xi_p,
\end{equation}
which are identified as the position coordinates
of $\x_1$ and $\x_2$ in Fig. \ref{area-corda}.
According to Appendix A, the Bohr-Sommerfeld quantization rule
\eqref{Bohr} leads to the corresponding phases as
\begin{eqnarray}
&&
S_+(q_1+\eq/2)-S_-(q_1-\eq/2)-\ep q_1
=\mathcal{A}_1(\xi)+\x_1\wedge\xi,
\\
&&
S_+(q_2+\eq/2)-S_-(q_2-\eq/2)-\ep q_2
=
\mathcal{A}_2(\xi)+\x_2\wedge\xi.
\end{eqnarray}
\par
Instead of evaluating \eqref{chiKB} by stationary phase, this integral is
mapped onto the standard form for a {\it fold diffraction catastrophe} \cite{Berry76}:
\begin{equation}
\chi(\xi)=
\frac{N^2}{2\pi i\hbar}
\exp\left(\frac{i\Sigma\mathcal{A}}{2\hbar}\right)
\int_{-\infty}^{\infty} du
\; g(u;\xi)\; e^{i\left[\frac{u^3}3-\zeta\,u\right]},
\label{chiairy}
\end{equation}
where we have defined
\begin{equation}
\label{Coef trans Corda}
\Sigma\mathcal{A}=\mathcal{A}_1+\mathcal{A}_2,
\quad\quad
\frac23\zeta(\Vxi)^{\frac32}
=
\frac{\mathcal{A}_{12}}{2\hbar}
=\frac{\mathcal{A}_{1}-\mathcal{A}_2+\Veta\wedge\Vxi}{2\hbar}.
\end{equation}
The action difference, $\mathcal{A}_{12}$, is the main ingredient
in the present application of the method of uniform approximation \cite{Chester, Berry76}.
Its geometric definition is the area between the closed curve and its translation,
as shown in Fig. \ref{area-corda}. At the caustic, the chord $\Vxi=\Vxi_D$
is maximal and its {\it conjugate chord} \cite{OzVaSa}:
$\eta=\x_1-\x_2\rightarrow 0$.

The above integral would define an Airy function
\cite{abramowitz}, Ai$(\zeta)$, if $g(u;\xi)=1$, but here the
mapping between the variables $q\mapsto u$, respectively in
\eqref{chiKB} and \eqref{chiairy}, leads to
\begin{equation}
g(u;\xi)=\left|\parcial{I}{p}(q(u)+\eq/2)\parcial{I}{p}(q(u)-\eq/2)\right|^{-\frac12}
\label{g-chord}
\deri{q}{u}.
\end{equation}
The approximation now consists in replacing this by a linear function
which coincides with $g(u;\xi)$ at the stationary points, $\pm\zeta^{\frac12}$ of \eqref{chiairy}.
These  map onto $q(\zeta^{\frac12})=\x_{1q}\equiv q_1$ and $q(-\zeta^{\frac12})=\x_{2q}\equiv q_2$.
The Jacobian of the mapping $q\leftrightarrow u$ at the stationary points
is specified by
\begin{eqnarray}
\deri{q}{u}(\zeta^{\frac12})&=&
\left|\frac{2\hbar\zeta^{\frac12}}
{\parcial{p}{q}(q_1+\eq/2)-\parcial{p}{q}(q_1-\eq/2)}\right|^{\frac12},\\
\deri{q}{u}(-\zeta^{\frac12})&=&\left|\frac{2\hbar\zeta^{\frac12}}
{\parcial{p}{q}(q_2+\eq/2)-\parcial{p}{q}(q_{2}-\eq/2)}\right|^{\frac12},
\end{eqnarray}
so that
\begin{equation}
g(\zeta^{\frac12};\xi)=\left|\frac{2\hbar\zeta^{\frac12}}{\{I^+_1,I^-_1\}}\right|^{\frac12},
\quad
g(-\zeta^{\frac12};\xi)=\left|\frac{2\hbar\zeta^{\frac12}}{\{I^+_2,I^-_2\}}\right|^{\frac12}.
\end{equation}
Thus, recalling the definition of the transported action \eqref{transaction1}
for each chord realization, we now define
\be
\Delta I^{12}\equiv
\left|\{I_1^+,I_1^-\}\right|^{-\frac12}-
\left|\{I_2^+,I_2^-\}\right|^{-\frac12}
\quad\textrm{and }\quad
\Sigma I^{12}\equiv
\left|\{I_1^+,I_1^-\}\right|^{-\frac12}+
\left|\{I_2^+,I_2^-\}\right|^{-\frac12},
\ee
so as to obtain
\begin{eqnarray}
g(u;\xi)&\simeq
\sqrt{\frac\hbar2}\left(\zeta^{\frac14}
\Sigma I^{12}
+
\frac{u}{\zeta^{\frac14}}
\Delta I^{12}\right),
\end{eqnarray}
as an approximation to \eqref{g-chord}, which is linear with respect to $u$
and has the correct values at the stationary points.
Thus, the integral for the chord function becomes
\begin{eqnarray}
\label{X-DOBRA}
\chi(\xi)&=&
\frac{N^2}{ih}\frac{\sqrt{\hbar}}{\sqrt2}
\exp\left(
          i\frac{\Sigma\mathcal{A}}{2\hbar}
     \right)
\int_{-\infty}^{\infty}
   du\,
   \left(
         \zeta^{\frac14}\Sigma I^{12}
         +
         \frac{u}{\zeta^{\frac14}}
         \Delta I^{12}
   \right)
   \,
   e^{i\left[
             \frac{u^3}3-\zeta\,u\right]}\nonumber\\
&=&
\frac{N^2}{i}\frac{1}{\sqrt{2\hbar}}
\exp\left(i\frac{\Sigma\mathcal{A}}{2\hbar}\right)
\,
\left(
      \zeta^{\frac14}\Sigma I^{12}
      \Ai\left(-\zeta\right)
      -
      \frac{i}{\zeta^{\frac14}}
      \Delta I^{12} \Ai'\left(-\zeta\right),
\right)
\end{eqnarray}
where $\Ai'$ is the derivative of the Airy function. 
Finally, recalling the definition of the intermediate variable,
$\zeta$, in terms of areas \eqref{Coef trans Corda}, 
we obtain the full unitary approximation for the chord function:
\begin{equation}
\chi(\Vxi)=N^2
\exp\left(i\frac{\Sigma\mathcal{A}}{2\hbar}\right)
\,
\left[
      \frac{\left[\frac{3}{4}\mathcal{A}_{12}\right]^{\frac16}\Sigma I^{12}}
      {i\sqrt{2}\hbar^{\frac23}}
      \Ai\left(-\left[\frac{3\mathcal{A}_{12}}{4\hbar}\right]^{\frac23}\right)\right.
      -\left.
      \frac{\left[\frac{3}{4}\mathcal{A}_{12}\right]^{-\frac16}\Delta I^{12}}
      {\sqrt{2}\hbar^{\frac13}}
      \Ai'\left(-\left[\frac{3\mathcal{A}_{12}}{4\hbar}\right]^{\frac23}\right)
\right]. 
\label{chordfold} 
\end{equation} 
\par
The Airy function,
$\Ai\left(-\zeta\right)$, oscillates with increasing amplitude as
its argument increases and then decays exponentially for positive
values of $-\zeta$. The maximum amplitude, just bellow the origin,
indicates the singularity of the simple semiclassical amplitude at
the caustic. In this region the second term depending on
$\Ai'\left(-\zeta\right)$ can be neglected, but it is necessary to
obtain the correct limiting behaviour in the oscillatory region
where the simple semiclassical description is valid. This term is
a new feature in comparison with the uniform approximation for the
Wigner function for small chords \cite{Berry77}, where it is
absent because of the reflection symmetry. The other novel feature
is the oscillatory phase proportional to $\Sigma\mathcal{A}$ along 
the caustic, which will be important to separate the
contribution of each realization at the oscillatory regime. Indeed
in the case when the chord areas are significantly
greater than Planck's constant, the functions in \eqref{chordfold}
can be replaced by the asymptotic forms for large negative
values \cite{abramowitz}, 
\be 
 \Ai(-x)\to\frac{1}{\sqrt\pi
x^{\frac14}}\cos\left(\frac23x^{\frac32}-\frac\pi4\right)\quad \textrm{and
\;}\quad
\Ai'(-x)\to\frac{x^{\frac14}}{\sqrt\pi}\sin\left(\frac23x^{\frac32}-\frac\pi4\right),
\ee 
in order to obtain the correct form in the oscillatory regime
as in \eqref{semichord}:
\be 
\chi(\xi)\to \frac{N^2}{i\sqrt{2\pi\hbar}}
\left[
      \frac{\exp\left(i\left[\frac{\mathcal{A}_{1}}{\hbar}-\frac\pi4
                                          \right]
                                    \right)}{ |\{I^+_1,I^-_1\}|^{\frac12}}
      +\frac{\exp\left(i\left[\frac{\mathcal{A}_{2}}{\hbar}+\frac\pi4
                                          \right]
                                    \right)}{|\{I^+_2,I^-_2\}|^{\frac12}}
\right].
\label{asympchord}
\ee
This result corresponds to the sum of contributions for each chord realization in the
simpler stationary phase approximation of the semiclassical
chord function in \cite{OzVaSa}, where it was considered that each chord realization
lies on a single of the WKB wave function.

%%%%%%
\section{Uniform approximation for the Wigner function}

\begin{figure}[t!]
\centering
\includegraphics[width=5cm]{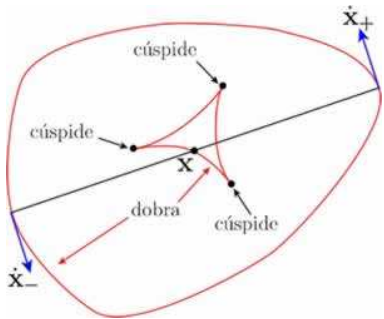}
\caption{Wigner function catastrophes for a closed curve. 
The curve itself is a fold ({\it short-chord catastrophe}). 
Another singular curve lying in the interior is composed of folds and cusps 
({\it long-chord catastrophe}). The catastrophe points correspond 
to centres of the diameters of the classical curve.
\label{catasWig}}
\end{figure}

As the Wigner function is evaluated at a point $\x$ that lies further and further
inside a convex closed classical curve, a caustic will be crossed. This is typically
a cusped triangle\cite{Berry77}, in which there are three chords for each centre, as
shown in Fig. \ref{catasWig}. Generic paths to its interior enter the triangle through
a fold caustic joining two cusps. Such a caustic point $\{\x_i\}$ is the centre of a
diameter, as show in Fig. \ref{catasWig}, as well being the centre of the other chord,
which was followed in the continuous path through the caustic. An example of the full 
fringe pattern where the regions characterized by one or three chords are clearly discernible
can be found in \cite{Oz-enta}.

\begin{figure}[t!]
\centering
\includegraphics[width=14cm]{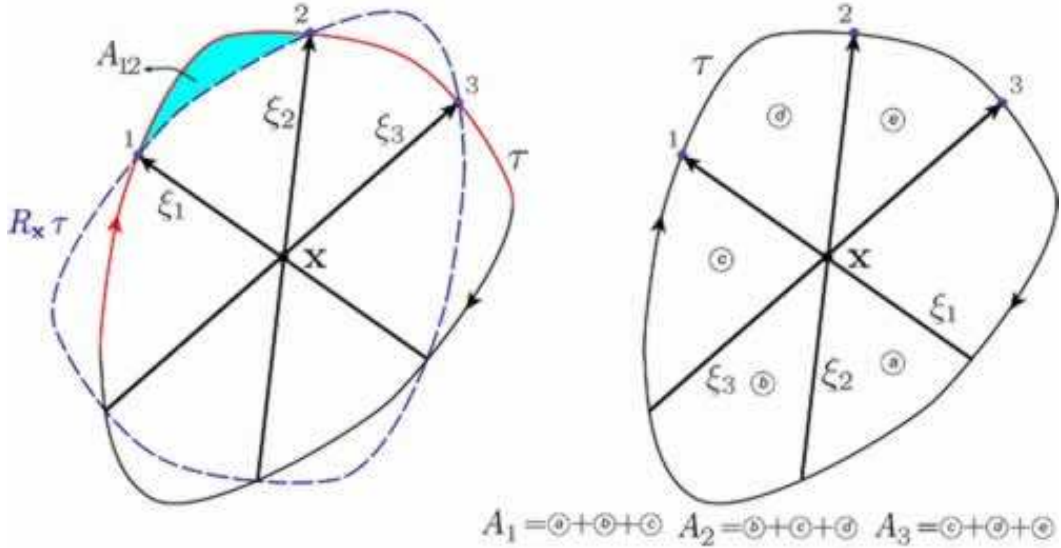}
\caption{Stationary chords near to long-chord catastrophe of the Wigner function, 
displaying the compositions of the area $A_j$ for each stationary chord. 
The phase difference for coalescent $\Vxi_1$ and $\Vxi_2$ chords 
is the symplectic area $A_{12}$ limited by their tips.
\label{dob-wig-cl}}
\end{figure}

The uniform approximation to be derived here concerns the pair of chords, 
$\Vxi_1$ and $\Vxi_2$ in Fig. \ref{dob-wig-cl}, 
that are born from the diameter at the caustic. 
Away from the cusps, the contribution of the separate chord, 
$\Vxi_3$ in Fig. \ref{dob-wig-cl}, can still be evaluated by stationary phase. 
This is just a simple semiclassical contribution,
\be
W_3(\x)=
\frac{4N^2}{\sqrt h}
\left|\{I(\x+\Vxi_{3}/2),I(\x-\Vxi_{3}/2)\}\right|^{-\frac12}
\cos\left(\frac{A_3}\hbar-\frac\pi4
\right),
\label{W_3}
\ee
which could be obtained from a single WKB branch, $i=j$ in \eqref{WigKB}. 
This could also be derived from a cross-branch $i\neq j$ 
(by rotating the phase space coordinates) 
from the same integral that furnishes the joint contribution of the pair of chords, 
$\Vxi_1$ and $\Vxi_2$, which coalesce at the diameter $\Vxi_D$. 

This crossed chord picture is essential for the uniform approximation. 
As in the theory for the chord function, 
we now map the integral in the region corresponding to the chords $\Vxi_1$ and $\Vxi_2$ 
onto the diffraction integral, as in \eqref{chiairy}.
\begin{equation}
W(\x)=2\Re\mathfrak{e}\frac{N^2}{i2\pi\hbar}{\exp\left(i\frac{\Sigma A}{2\hbar}\right)}
\int_{-\infty}^\infty dz\, g(z;\x)\,e^{i\left[\frac{z^3}{3}-\zeta z\right]}+W_3(\x).
\label{W-diffra}
\end{equation}
Here the parameters of the transformation are given by
\begin{equation}
\Sigma A=A_1(\x)+A_2(\x),\quad\quad
\frac23\zeta^{\frac32}(\x)=\frac{A_{12}(\x)}{2\hbar}=\frac{A_1(\x)-A_2(\x)}{2\hbar},
\end{equation}
in which $A_{12}$ is the symplectic area bounded by the closed curve 
and its reflection between the ends of $\Vxi_1$ and $\Vxi_2$, as show Fig. \ref{dob-wig-cl}. 
The procedure is straightforward as for the chord function, 
so that recalling that $I_j^{\pm}(\x)=I(\x\pm\xi_j/2)$, we now define
\be
\Delta I_{12}\equiv
\left|\{I_1^+,I^-_1\}\right|^{-\frac12}-
\left|\{I_2^+,I_2^-\}\right|^{-\frac12}
\quad
\textrm{and}
\quad
\Sigma I_{12}\equiv
\left|\{I_1^+,I_1^-\}\right|^{-\frac12}+
\left|\{I_2^+,I_2^-\}\right|^{-\frac12}.
\ee
Hence, the linear approximation for the amplitude in \eqref{W-diffra} is
\be
g(z,\x)=
\sqrt{2\hbar}\left(
\zeta^{\frac14}\Sigma I_{12}
+\frac{z}{\zeta^{\frac14}}\Delta I_{12}
\right),
\ee
so that, the Wigner function is given by
\begin{eqnarray}
W(\x)
=
2\sqrt2N^2
\left[
      \frac{\left[
                 \frac{3}{4} A_{12}
            \right]^{\frac16}\,\Sigma I_{12}
            \sin\left(
                     \frac{\Sigma A}{2\hbar}
                \right)
           }{
             \hbar^{\frac23}
            }
      \Ai\left(
               -\left[
                      \frac{3 A_{12}}{4\hbar}
                \right]^{\frac23}
         \right)
       \right.\nonumber\\
\hspace{2cm}\left.-
       \frac{\Delta I_{12}
             \cos\left(
                       \frac{\Sigma A}{2\hbar}
                 \right)
            }{
             \hbar^{\frac13}\left[
                                  \frac{3}{4} A_{12}
                            \right]^{\frac16}
            }
       \Ai'\left(
                 -\left[
                        \frac{3 A_{12}}{4\hbar}
                  \right]^{\frac23}
           \right)
\right]+W_3(\x).
\end{eqnarray}

Again, as for chord function, the behaviour near the caustic is described correctly in terms of the Airy function and its derivative. The fact that the Wigner function also has cusps, which are of higher order than the catastrophes of the chord function, adds an additional term. Crossing the fold caustic, the coalescent chords disappear and only the additional term remains, coinciding with the simpler stationary phase approximation far form the caustic\cite{Berry77}, although the normalization constant must be re-evaluated.
\par For regions where $A_{12}\gg \hbar$, we can replace the Airy function and its derivative by their asymptotic forms for large negative values, obtaining
\begin{eqnarray}
W(\x)
=
\frac{4N^2}{\sqrt{2\pi\hbar}}
\left[
\frac{\sin\left(\frac{A_1}{\hbar}-\frac{\pi}4
          \right)
     }
     {
|\{I_1^+,I_1^-\}|^{\frac12}}
+
\frac{\sin\left(\frac{A_2}{\hbar}+\frac{\pi}4
          \right)
     }
     {|\{I_2^+,I_2^-\}|^{\frac12}}
+
\frac{\cos\left(\frac{A_3}{\hbar}-\frac\pi4
      \right)
     }
     {|\{I_3^+,I_3^-\}|^{\frac12}}
\right]
\label{asympW}
\end{eqnarray}
which is a sum of oscillatory terms, each one with a different phase, 
as in \eqref{semiwigner}. This asymptotic form is a superposition 
of the individual stationary phase approximations for each stationary chord.

%-----------------Transicional-----------------------------------------------------------

\section{Approximations for the transitional regions}

The uniform approximation \eqref{chordfold} is not explicitly resolved 
very close to the caustic, because $\mathcal{A}_{12}\to0$ as the caustic is approached, 
whereas $\Sigma I^{12}\to\infty$. The classical curve can be approximated by parabolae 
in both the neighborhoods of the tips of the realization of $\Vxi_D$. 
This equates the amplitude associated for each stationary point, 
i.e. $\Sigma I^{12}=2\{I^+,I^-\}$ and $\Delta I^{12}\to0$. 
Thus, the term of the derivative of the Airy function in \eqref{chordfold} 
cancels near to the caustic.
\par
\begin{figure}
\centering
\includegraphics[width=6cm]{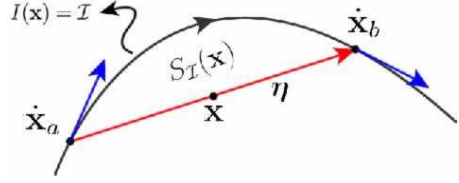}
\caption{The area $S_{\mathcal{I}}$ for a point very close to the level curve, 
$I(\x)=\mathcal{I}$. The tips of the chord $\Veta$ evolve 
by the action of the Hamiltonian $I(\x)$. 
As a first approximation, this evolution is linear.
\label{trans}}
\end{figure}

To obtain an explicit expression for the transitional chord function, 
we start by recalling that the action variable, $I(\x)$, 
can be interpreted as a Hamiltonian, such that the classical curve is a trajectory, 
i.e. the level curve $I(\x)=\mathcal{I}$. 
Considering $\x$ as the centre of a chord $\Veta$ that conects two points of the curve, 
$\x_a$ and $\x_b$, we obtain as a first approximation, $\x_b\simeq\x_a+\tau\dot{\x}_a$,
if $\x$ is very close to this curve. Then the action can be expanded as 
(see Appendix B in Ref. \cite{OzorioRep})
\be
\mathcal{I}-I(\x)\simeq\frac1{8}\tau^2\,\dot\x\,\mathfrak{I}_\x\,\dot\x,
\label{I-I(x)}
\ee
where $\mathfrak{I}_\x$ is the Hessian matrix of $I$ at the point $\x$. 
This quadratic Hamiltonian generates the assumed linear motion. 
On the other hand, the area between $\Veta$ and the curve (Fig. \ref{trans})
is given by \cite{OzorioRep}.
\be
S_{\mathcal{I}}(\x)\simeq\frac{1}{12}\tau^3\,\dot\x\,\mathfrak{I}_\x\,\dot\x.
\ee
Thus, noticing that the centres of the realizations of $\Veta$ are $\X^\pm\equiv\pm\Vxi/2$, 
the symplectic area $\mathcal{A}_{12}$ in Fig. \ref{area-corda} may be obtained as
\be
\mathcal{A}_{12}(\Vxi)=S_{\mathcal{I}}(\Vxi/2)+S_{\mathcal{I}}(-\Vxi/2)=
\frac1{12}\left(
                \tau_+^3\dot{\X}^+\mathfrak{I}_{\frac{\bm{\xi}}2}\dot{\X}^+
               +\tau_-^3\dot{\X}^-\mathfrak{I}_{\scalebox{1}{-}\frac{\bm{\xi}}2}\dot{\X}^-
          \right)
\label{A12 trans I}
\ee
where $\tau_\pm$ is the time of flight between the tips 
of each realization of $\Veta$, under the action of the Hamiltonian $I(\x)$.
\par
Recalling that $\x_1$ and $\x_2$ are the midpoints of the realizations of $\Vxi$, 
the Poisson brackets of the action in the amplitudes in \eqref{chordfold} 
will be given by the symplectic products, 
\be
\dot\x_1^+\wedge\dot\x_1^-=\{I_1^+,I_1^-\}\simeq-\{I_2^+,I_2^-\}=\dot\x_2^-\wedge\dot\x_2^+
\ee
So that defining the `accelerations'\cite{OzorioRep,Berry89}, $\ddot\x_1$ and $\ddot\x_2$, as
\be
\ddot\x_j=\left(\dot\x_j\cdot
 \frac{\partial}{\partial\x}
\right)\dot\x=\J\mathfrak{I}_\x\dot\x,
\ee
such that $\dot\x_2^+\simeq\dot\x_1^++\tau_+\ddot\x_1^+$ 
and $\dot\x_1^-\simeq\dot\x_2^-+\tau_-\ddot\x_2^-$, we obtain
\be
\{I_1^+,I_1^-\}
=
 \frac12\left[
              \tau_+\dot\X^-\mathfrak{I}_{\frac\xi2}\dot\X^++
              \tau_-\dot\X^+\mathfrak{I}_{\scalebox{1}{-}\frac\xi2}\dot\X^-
        \right]
\ee 
The times $\tau_+$ and $\tau_-$ can be found using \eqref{I-I(x)}, 
so that \eqref{A12 trans I} becomes 
\be
\frac34\mathcal{A}_{12}(\Vxi)=
\sqrt2\left[
            \frac{[\mathcal{I}-I(\frac{\bm{\xi}}2)]^{\frac32}}
                 {[\dot\X^+\mathfrak{I}_{\frac{\bm{\xi}}2}\dot\X^+]^{\frac12}}
           +\frac{[\mathcal{I}-I(\scalebox{1}{-}\frac{\bm{\xi}}2)]^{\frac32}}
                 {[\dot\X^-\mathfrak{I}_{\scalebox{1}{-}\frac{\bm{\xi}}2}\dot\X^-]^{\frac12}}
           \right].
\label{A12 trans}
\ee
Thus, we have all the necessary ingredients to obtain the transitional form of the chord function, 
\begin{eqnarray}
\chi(\Vxi)&=&\frac{e^{i\Sigma\mathcal{A}/2\hbar-i\pi/2}}{\pi^{\frac13}h^{\frac23}}
\frac{[\tau_+^3\dot\X^+\mathfrak{I}_{\frac{\bm{\xi}}2}\dot\X^+
      +\tau_-^3\dot\X^-\mathfrak{I}_{\scalebox{1}{-}\frac{\bm{\xi}}2}\dot\X^-]^{\frac16}}
     {[\tau_+\dot\X^-\mathfrak{I}_{\frac{\bm{\xi}}2}\dot\X^+
      +\tau_-\dot\X^+\mathfrak{I}_{\scalebox{1}{-}\frac{\bm{\xi}}2}\dot\X^-]^{\frac12}}
\nonumber\\
&&\hspace{2.5cm}\times
\Ai\left(
         -2^{\frac13}
           \left[\frac{[\mathcal{I}-I(\frac{\bm{\xi}}2)]^{\frac32}}
                      {[\dot\X^+\mathfrak{I}_{\frac{\bm{\xi}}2}\dot\X^+]^{\frac12}}+
                 \frac{[\mathcal{I}-I({\scalebox{1}{-}\frac{\bm{\xi}}2})]^{\frac32}}
                      {[\dot\X^-\mathfrak{I}_{\scalebox{1}{-}\frac{\bm{\xi}}2}\dot\X^-]^{\frac12}}
           \right]^{\frac23}
   \right).
\label{trans-chord}
\end{eqnarray}

If the curve has a local symmetry of reflection with respect to the origin, 
the hessian matrices will be equal, 
i.e. $\mathfrak{I}_{\frac{\bm{\xi}}2}=\mathfrak{I}_{\scalebox{1}{-}\frac{\bm{\xi}}2}$ 
and the velocity vectors  $\dot\X^+=-\dot\X^-$. 
Here we recall that the origin depends of the chord $\Vxi$, 
since it has been chosen to be the midpoint between the centres of its realizations 
on the closed curve. Thus, the transitional chord function reduces in this simple case to
\be
\chi(\Vxi)=\frac{e^{i\Sigma\mathcal{A}/2\hbar+i\pi}}{(2\pi)^{\frac13}h^{\frac23}}
[\dot\X^+\mathfrak{I}_{\frac{\bm{\xi}}2}\dot\X^+]^{-\frac13}
\Ai\left(
         2\frac{I(\frac{\bm{\xi}}2)-\mathcal{I}}
                      {[\dot\X^+\mathfrak{I}_{\frac{\bm{\xi}}2}\dot\X^+]^{\frac13}}
   \right).
\label{transchi}
\ee
For a diameter $\Vxi_D$, the argument of the Airy function cancels, 
because $I(\pm\Vxi_D/2)=\mathcal{I}$, and $\Sigma\mathcal{A}/2$ is the chord area of $\Vxi_D$. 
Thus, the transitional approximation remains finite, and at the caustic it is close 
to a local amplitude maximum.
\par
We can follow a similar procedure for the Wigner function. 
Defining $\bar{\Vxi}$, as the average between the stationary chords $\Vxi_1$ and $\Vxi_2$, 
together with the pair of phase space points, $\bm{y}^\pm=\x\pm\bar{\Vxi}/2$, we obtain
\begin{eqnarray}
W(\Vxi)&=&\frac{4\sin\left[\frac{\Sigma\mathcal{A}}{2\hbar}\right]}{\pi^{\frac13}h^{\frac23}}
\frac{[\tau_+^3\dot{\bm{y}}^+\mathfrak{I}_{{\bm{y}}^+}\dot{\bm{y}}^+
      -\tau_-^3\dot{\bm{y}}^-\mathfrak{I}_{{\bm{y}}^-}\dot{\bm{y}}^-]^{\frac16}}
     {[\tau_-\dot{\bm{y}}^+\mathfrak{I}_{{\bm{y}}^-}\dot{\bm{y}}^-
      -\tau_+\dot{\bm{y}}^-\mathfrak{I}_{{\bm{y}}^+}\dot{\bm{y}}^+]^{\frac12}}
\nonumber\\
&&\hspace{2.5cm}\times
\Ai\left(
         -2^{\frac13}
           \left[\frac{[\mathcal{I}-I(\bm{y}^+)]^{\frac32}}
                      {[\dot{\bm{y}}^+\mathfrak{I}_{\bm{y}^+}\dot{\bm{y}}^+]^{\frac12}}-
                 \frac{[\mathcal{I}-I(\bm{y}^-)]^{\frac32}}
                      {[\dot{\bm{y}}^-\mathfrak{I}_{\bm{y}^-}\dot{\bm{y}}^-]^{\frac12}}
           \right]^{\frac23}
   \right).
\label{trans-wig}
\end{eqnarray}
Unlike the chord function, this transitional form is undetermined at the caustic,
in the particular case that the state is invariant with respect to a reflection symmetry,
because then the entire caustic colapses to a point, where the Wigner function 
has a catastrophe of higher order.

%%-------Fock states--------------------------------------

\section{Long-chord regime for Fock states}

Now we consider the excited states, $|n\rangle$, of the one dimensional harmonic
oscillator whose classical manifold is a circumference centred at the origin, 
i.e. it has reflection symmetry. The
quantization condition \eqref{Bohr} for these circles defines 
\be
\pi(p^2+q^2)=
2\pi\hbar\left(n+\frac12\right). 
\ee

The exact chord function is given by \cite{OzVaSa} 
\be
\chi_n(\xi)=\frac{e^{-\xi^2/4\hbar}}{2\pi\hbar}L_n\left(\frac{\xi^2}{2\hbar}\right),
\ee where $L_n$ is a Laguerre polynomial. For small chords, $\xi\ll\hbar$,
\be
\chi_{\mathcal{I}}(\xi)\simeq\frac1{2\pi\hbar}J_0\left(\frac{\sqrt{2\mathcal{I}}|\xi|}\hbar\right),
\label{besselchord} 
\ee
gives a good approximation for the chord function \cite{OzVaSa}, where $J_0$,
is the Bessel function of order zero. Then, the asymptotic form of the Bessel
function for large values \cite{abramowitz} leads to \be
\chi_{\mathcal{I}}(\xi)\simeq
\frac{(\sqrt{2\mathcal{I}}|\xi|)^{-\frac12}}{\pi\sqrt{h}}
\cos\left(\frac{\sqrt{2\mathcal{I}}|\xi|}{\hbar}-\frac{3\pi}4\right).
\ee
\par 
We can compare this result with the oscillatory regime \eqref{asympchord}. 
First, due to symmetry, the area of one chord realization 
is complementary to the area of the other and a simple integral gives the semiclassical phase as
\be
\mathcal{A_I}(\xi)=
2\pi\mathcal{I}-|\xi|\sqrt{2\mathcal{I}-\left(\frac{|\xi|}{2}\right)^2}
-4\mathcal{I}\arcsin\left(\frac{|\xi|/2}{\sqrt{2\mathcal{I}}}\right).
\label{A fock}
\ee
Moreover, symmetry equates the Poisson bracket for each chord realization, 
so that terms containing the derivative of Airy function will cancel. 
The Poisson brackets are then evaluated as
\be
\{I^+,I^-\}=|\xi|\sqrt{2\mathcal{I}-\left(\frac{|\xi|}2\right)^2},
\ee
so that, the uniform approximation for the chord function becomes
\be
\chi_{\mathcal{I}}(\xi)=
(-1)^n\frac{\sqrt2N^2\left[\frac34\mathcal{A_I}(\xi)
               \right]^{\frac16}}
     {\hbar^{\frac23}|\xi|^{\frac12}}
\left[2\mathcal{I}-\frac{|\xi|^2}4
\right]^{-\frac14}
\Ai\left(-\left[\frac{3\mathcal{A_I}}{4\hbar}
          \right]^{\frac23}
   \right).
\ee
It follows that the asymptotic behaviour of Airy function, extrapolated for small chords is
\be
\chi_{\mathcal{I}}(\xi)=
(-1)^n
\frac{2N^2}{\sqrt h}
\left(\sqrt{2\mathcal{I}}|\xi|
\right)^{-\frac12}
\cos\left(\frac{\mathcal{A_I}}{2\hbar}-\frac{\pi}4
     \right).
\ee 
Note that, to lowest order, the argument in the
Bessel function \eqref{besselchord}, $\sqrt{2\mathcal{I}}|\xi|$, is
one half of the complementary area to the intersection between the
circle and its translation. 

Thus, the asymptotic limit of the
chord function for the long-chord caustic reproduces the chord
function of small chords for Fock states, in an intermediary
region. Furthermore, we immediately obtain the normalization constant as 
\be
N^2=\frac1{2\pi}.
\ee 
This is an alternative derivation to Berry's \cite{Berry77}. 
We can replace this value in \eqref{asympW} outside the caustic,
where the two coalescent chords disappear, 
so as to recover the simpler stationary phase approximation
for the Wigner function \cite{Berry77}.

%--------------Final remarks ---------------------------------
\section{Discussion}

We have shown that the behaviour of both the Wigner function and the chord function near a
maximal chord singularity can be described by the Airy function and its
derivative. Although, the latter becomes negligible for points very close to the caustic, 
it adds an important contribution to the expansion of the semiclassical distributions 
in the oscillatory region, coinciding there with the simpler stationary phase method. 
The shape of the diameter-caustic is different in the phase space of centres, 
$\x$, where the Wigner function is defined, and in the space of chords, $\Vxi$.
In the latter case, the caustic is located on the locus of diameters, $\Vxi_D$, 
maximal chords of the original closed complex curve. This diameter caustic is symmetrical 
with respect to the chord origin and it is also convex. If the assumption of
convexity is relaxed, the symmetry will be preserved, 
because $-\Vxi_D$ is also a diameter, but the simple fold caustic 
may then exhibit higher singularities. This is the case for the diameter-caustic
viewed in the phase space of centres $\x_D=\x(\Vxi_D)$. 
This caustic of the Wigner function has cusps even 
in the case of a convex quantized curve \cite{Berry77}.

The pair of caustics that concern us may be viewed as alternative projection singularities 
of a single (lagrangian) surface in a doubled phase space, $\X=\x^-\times\x^+$, 
i.e the product torus for the pair of quantized curves $I(\x^{\pm})=\cal{I}$. 
The alternative description of double phase space in terms of the centres, 
$\x=(\x^+ + \x^-)/2$ and $\Vxi=\Vxi^+ - \Vxi^-$ leads to a description of the
double torus that no longer factors \cite{Oz-enta}. The points $\X_D=(\x(\Vxi_D), \Vxi_D)$ 
on the double torus project singularly onto both these double phase space 
coordinate planes. Thus the two-dimensional tangent plane to the torus
is completely determined by the pair of tangent vectors: $(\x(\Vxi_D), 0)$
and $(0, \Vxi_D)$.
\footnote{It is an unfortunate confusion that the canonical variable for
double phase space is $\y=\J\vxi$(where $\J$ is the symplectic matrix (\ref{eqc.4}))
as in \cite{Oz-enta}, but this is not important for the present discussion.}
 
\par
Let us summarize the behaviour of the chord function for any translation: 
$a)$ a maximum at the origin; $b)$ an oscillatory regime, 
obtained as a superposition of stationary phase terms for each chord realization; 
$c)$ a region near to the maximal chord, i.e. the diameter $\Vxi_D$, 
expressed in terms of the Airy function and its derivative, 
where the amplitude is again maximal and finally 
$d)$ an evanescent region for chords longer than diameters
(also described by the Airy functions).

We have considered only pure states, for which the phase space
correlation \eqref{pscor} is given by the square modulus of the
chord function. Thus, using the asymptotic form of the uniform
approximation for the chord function at the semiclassical regime,
we find that, in the oscillatory region far from the caustic, the phase
space correlation is approximately
\be 
C_\xi= \frac{1}{\hbar} \left[
\{I_1^+,I_1^-\}^{-1}+\{I_2^+,I_2^-\}^{-1}
+2\{I_1^+,I_1^-\}^{-\frac12}\{I_2^+,I_2^-\}^{-\frac12}
\sin\left(\frac{\mathcal{A}_{12}}{\hbar}\right) \right],
\label{semi-corr} 
\ee 
i.e. a pair of classical terms associated to each chord realization and 
a term that represents their interference. This formula corrects the semiclassical 
phase space correlation presented in \cite{OzVaSa}, 
which also provides a semiclassical interpretation for the invariance of the correlation 
with respect to Fourier transformation. The important point is that the
maximal reach of the phase space correlations (also the correlations of the Wigner function) 
in the neighbourhood of a diameter of the 
quantized curve is just $C_\xi=|\chi(\Vxi)|^2$ in the simple form given by \eqref{transchi}.

\section*{ACKNOWLEDGMENTS}
Partial financial support from, CNPq, FAPERJ and CAPES (brazilian agencies),
as well as UNESCO/IBSP Project 3-BR-06 is gratefully acknowledged.

\appendix

\section{Crossed chords}

\begin{figure}[t!]
\centering
\includegraphics[width=11cm]{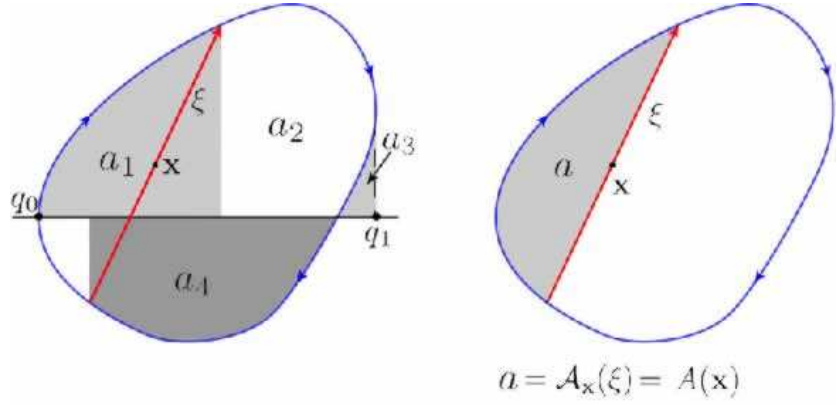}
\caption{Phase difference between the stationary phase points for the Wigner function,
evaluated at $\x=(p,q)$, such that the tips of its chord, $\xi=(\ep,\eq)$, lie on different
branches of the WKB function. The shaded area, $a=S_+(q+\xi_q/2)-S_-(q-\eq/2)-p\eq$,}
\label{AP-fase}
\end{figure}

Redefining $S_i=S_+$ and $S_j=S_-$ in \eqref{WigKB}, the phase in the integrand 
for the Wigner function evaluated at the point $\x=(p,q)$ is
\begin{equation}
\hbar F(Q;\x)=S_+(q+Q/2,I)-S_-(q-Q/2)-pQ.
\end{equation}
Let $q_0$ and $q_1$ be the turning points on the closed curve, with $q_0<q_1$.
Choosing the $q$-axis to pass through $q_0$, according to Fig. \ref{AP-fase}, 
then at the stationary phase points, $Q=\pm \Vxi_q$, we have 
\begin{eqnarray}
 S_+(q+\xi_q/2)&=&
\int_{q_0}^{q+\eq/2}p_+(Q)dQ=a_1,
\\
 S_-(q-\xi_q/2)&=&\int_{q_0}^{q_1}p_+(Q)dQ+\int_{q_1}^{q-\xi_q/2}p_-(Q)dQ=a_1+a_2+a_3+[-a_3+a_4].
\end{eqnarray}
The stationary chord, $\xi$, is here assumed to be crossed,
i.e to have both its tips on different action branches. Then, the phase becomes
\be
\hbar F(\xi_q;\x)=a_1-[a_1+a_2+a_4]-p\eq=-[a_2+a_4]-p\eq=a-\oint p dq.
\ee
Defining $A(\x)$ as the area between the stationary chord and the closed curve, leads to
\be
\hbar F(\xi_q;\x)=A(\x)-\oint p dq =A(\x)-2\pi\left(n+\frac12\right).
\label{wigner-ph}
\ee
%------------------------FASE - CORDAS-----------------------------------------------------------

On the other hand, the phase of the integrand in \eqref{chiKB} 
that defines the chord function for the same geometry is
\begin{equation}
\hbar F_\chi(Q,\xi)=S_+(Q+\xi_q/2)-S_-(Q-\xi_q/2)-\xi_p Q.
\end{equation}
The stationary points are position coordinates $q$, 
but their geometrical interpretation provides the respective momentum coordinates, 
$p$ (as discussed in sec. 2). Again we denote $\x=(p,q)$, thus the stationary phase $F_\chi$ depends of $\x$ instead of only $q$. Choosing the $q$-axis in the same way as before, we obtain
\begin{eqnarray}
\hbar F_\chi(\x,\xi)
&=S_+(q+\xi_q/2)-S_-(q-\xi_q/2)-p\eq+p\eq-\xi_p q,\\
&=S_+(q+\xi_q/2)-S_-(q-\xi_q/2)-p\eq+\x\wedge\xi
\end{eqnarray}
Defining $\mathcal{A}_\x(\xi)$ as the symplectic area between the closed curve
and the realization of  $\xi$, centred on $\x$, as in Fig. \ref{AP-fase}, leads to
\begin{equation}
\hbar F_\chi(\x,\xi)=\mathcal{A}_\x(\xi)+\x\wedge\xi-2\pi\left(n+\frac12\right).
\label{chord-ph}
\end{equation}
In order to implement the uniform approximations in sec. 3 and 4, we
can ignore the additional term $\oint pdq$ in \eqref{wigner-ph} and
\eqref{chord-ph}, because it is quantized. It should be
recalled that the quantization of the closed curve implies that
the areas, $A(\x)$ and $\mathcal{A}_\x(\xi)$ and their respective
complementary areas, $A'(\x)$ e $\mathcal{A}_\x'(\xi)$, satisfy
the rule
\begin{equation}
 A(\x)+A'(\x)=\mathcal{A}_\x(\xi)+\mathcal{A}_\x'(\xi)=\oint p\,dq = 2\pi\hbar\left(n+\frac12\right),\quad\quad\quad n\in\mathbb{Z}.
\end{equation}
%--------------------------------------REFERENCIAS-----------------------------------------------

%\newpage

\end{document}